\documentstyle{cupconf}

\input epsf
\epsfverbosetrue

\ifoldfss
\else
  \ifnfssone
    \newmathalphabet{\mathit}
      \addtoversion{normal}{\mathit}{cmr}{m}{it}
      \addtoversion{bold}{\mathit}{cmr}{bx}{it}
    \newmathalphabet{\mathcal}
      \addtoversion{normal}{\mathcal}{cmsy}{m}{n}
    \else
    \ifnfsstwo
    \fi
  \fi
\fi

\def\subsun{_\odot}

\def\hexnumber#1{\ifcase#1 0\or1\or2\or3\or4\or5\or6\or7\or8\or9\or
 A\or B\or C\or D\or E\or F\fi }

\makeatletter
\ifx\CUP@mtlplain@loaded\undefined
\else
\fi
\makeatother
 \makeatletter
 \ifx\CUP@mtlplain@loaded\undefined
   \font\tenbmi=cmmib10 at 10pt
   \font\sevenbmi=cmmib10 at 7pt
   \font\fivebmi=cmmib10 at 5pt

   \newfam\bmifam
   \textfont\bmifam=\tenbmi
   \scriptfont\bmifam=\sevenbmi
   \scriptscriptfont\bmifam=\fivebmi
   
 \fi
 \makeatother
\ifnfsstwo

\fi
\ifnfssone

\fi
\ifoldfss

\fi

\mathchardef\varLambda="0103

\makeatletter
\ifx\CUP@mtlplain@loaded\undefined
\else
\fi
\makeatother
\makeatletter
\ifx\CUP@mtlplain@loaded\undefined
  \font\tenbms=cmbsy10
  \font\sevenbms=cmbsy10 at 7pt
  \font\fivebms=cmbsy10 at 5pt
  \newfam\bmsfam
  \textfont\bmsfam=\tenbms
  \scriptfont\bmsfam=\sevenbms
  \scriptscriptfont\bmsfam=\fivebms

  \edef\bsy@{\hexnumber\bmsfam}
  \mathchardef\bnabla="0\bsy@72
\fi
\makeatother
\def\eg{{e.g.,\ }}

\def\etal{\mbox{\it et al.\ }}

\title[Planetary Nebula Luminosity Function]{Verifying the Planetary Nebula
Luminosity Function Method}

\author[G. H. Jacoby]%
{G\ls E\ls O\ls R\ls G\ls E\ns H.\ns J\ls A\ls C\ls O\ls B\ls Y$^1$}

\affiliation{$^1$National Optical Astronomy Observatories, P.O. Box 26732, Tucson, AZ 85726}

\begin{document}
\ifnfssone
\else
  \ifnfsstwo
  \else
    \ifoldfss
      \let\mathcal\cal
      \let\mathrm\rm
      \let\mathsf\sf
    \fi
  \fi
\fi

\maketitle

\begin{abstract}
The planetary nebula luminosity function (PNLF) technique for determining
distances to galaxies now has been applied to 34 galaxies, including
6 in the Virgo cluster and 3 in the Fornax cluster. Of these, 16 galaxies
are late-type or spirals and presumably contain Cepheid variables useful
for verifying the PNLF method. For 7 of these galaxies, Cepheid distances
exist; the PNLF distances agree with the Cepheid distances within the
dispersion of 8\% and within a zero-point offset of 1\%.

In addition, 3 small groups were studied (NGC 1023, Leo I, and Coma I)
where both spiral and elliptical distances were obtained to investigate
the magnitude of any systematic dependence on spiral versus elliptical
Hubble type.  None was found. Since the PNLF method agrees well with
the Cepheid system, and there is no measurable dependence on Hubble
type, it follows that PNLF distances to the ellipticals in Virgo and
Fornax also are on the Cepheid scale. This conclusion is strengthened
by the Cepheid distances to several Virgo galaxies and the recent
determination of a Cepheid distance to Fornax.

Challenges to the PNLF method by Bottinelli \etal (1991) and Tammann
(1992) are demonstrated to be incorrect. In particular, the allegation
that the Virgo distances suffer from inadequate survey depth is
rendered baseless with recent observations that extend the PNLF beyond
the power law regime and well into the plateau region. Using the new
observations of 320 PN, bootstrap tests show that any sample size
effect is smaller than 3\%.

Finally, a simple thought experiment is presented whereby M87 is placed
at 22 Mpc as argued by those favoring H$_0$ $\sim50$. The consequent
luminosities for observed planetary nebulae are inconsistent with
stellar evolution theory, thereby invalidating the assumption of a
distance greater than $\sim 17$ Mpc; alternatively, a drastic change in
stellar evolution theory is required.

\end{abstract}

\firstsection 
\section{Introduction}

The Planetary Nebula Luminosity Function (PNLF) technique for
determining distances to galaxies is described in \cite{Jacoby92}.
Briefly, the method relies on the constancy of the [OIII] $\lambda5007$
luminosity at which the population of high luminosity planetary nebulae
(PN) declines rapidly.  While the number of PN decreases monotonically
with increasing luminosity, there exists a luminosity beyond which PN
are not observed at all.  This limiting luminosity is a consequence of
several factors (Jacoby (1989), Kaler \& Jacoby (1991), M\'endez \etal
(1993), Han \etal (1994), Stanghellini (1995), Richer \etal (1997)), the
most important being the independence between progenitor
age (or initial mass) and the final white dwarf (or central star) mass
over reasonable age ranges (\eg 3--11 Gyrs).  For the great majority of
progenitors and therefore, for normal populations, the critical PN
luminosity varies by less than 0.1 mag.

The PNLF technique is one of the few that has been demonstrated to
yield consistent results in both elliptical and spiral galaxies at
distances up to 20 Mpc.  Consequently, it offers a path to unify the
population I distances (\eg Cepheids) with population II objects (\eg
elliptical galaxies). The former are needed for calibration while the
latter are needed to define distances to rich clusters.  Other
promising cross-population indicators include SN Ia when corrected for
decline rate variations (\cite{Hamuy96}), surface brightness
fluctuations (SBF), and globular cluster luminosity functions (GCLF).

This paper describes recent results in which the PNLF has been extended
to late-type galaxies in order to improve the zero-point calibration,
and new observations in M87 to assess the legitimacy of the arguments
posed against the use of the PNLF by \cite{Bottinelli91} and
\cite{Tammann92}.

\section{What Makes a Good Distance Indicator?}

Several speakers at this meeting presented their views on the important
qualities of a good distance indicator.  They noted the advantages of
an indicator that is luminous, easily identified, and has easily
measured properties (\eg magnitudes). My list focuses more
on the physical nature of the indicator rather than its technical
properties. Thus, a good distance indicator:

\medskip
\begin{enumerate}
\renewcommand{\theenumi}{\arabic{enumi}}
\item Has a good zero point calibration
\item Has a good prescription for metallicity correction
\item Has a good prescription to correct for effects of stellar ages
\item Can be corrected for effects of foreground and internal extinction
\item Has a good physical rationale
\item Can be tested (a) against other methods and (b) by other investigators
\end{enumerate}
\medskip

\noindent
This paper concentrates on the first of these. It is worth reviewing, though,
how well the PNLF satisfies the remaining 5 criteria.

Metallicity effects were modeled by \cite{Dopita92}. Limited
observational testing was performed by \cite{CJ92}. Most galaxies of
interest have metallicities within a factor of 2 of solar abundances,
and the predicted and observed errors in PNLF distances are smaller
than 5\% over this range.

The effects of population age on distance have been modeled by \cite{Dopita92},
\cite{Mendez93}, and \cite{Stanghellini95} and shown to be 
$<$5\% for galaxies having ages in the oberved range between 3 and 11
Gyrs.  Direct tests are complicated by our present inability to measure
population ages accurately, but PNLF distances to young (LMC, SMC,
M101) and old populations (M31's bulge, M81's bulge) where distances
are known from Cepheids, fail to detect any age effect at all.

Internal extinction is less of a problem than intuitively suspected.
In ellipticals, extinction is not an issue since the dust density is
very low. In spirals, significant errors are expected
if internal extinction in the galaxy is ignored. Observations, though,
fail to reveal any measurable distance errors in the 7 late-type
calibrators (Feldmeier, Ciardullo, \& Jacoby 1997). 
Feldmeier, Ciardullo, \& Jacoby (1997) modeled the
effects of dust to understand this unexpected situation, and
found that PN, which generally have scale heights well above the
population I disk of a spiral, are either so extincted by heavy dust
that they fall out of the PNLF sample, or they are so little affected
that their magnitudes are not significantly diminished. 

The excellent agreement between the PNLF and Cepheids (as well as SBF
and with other methods to a lesser degree) demands that a physical
basis for the PNLF must exist. Before all the comparisons were made,
though, the theory had been described by \cite{Jacoby89}, \cite{Dopita92},
and \cite{Mendez93}; recently, Stanghellini (1995) investigated
the H$\beta$ PNLF. In short, it has proven easy to reproduce the constancy
of the PNLF provided the population age is within the range of 3 to 11 Gyrs.
If the progenitors are as young as 0.5 Gyrs, the PNLF brightens by $\sim$0.3 mag.
And, if all progenitors in a galaxy are much older than 11 Gyrs, they fail to
produce observable PN at all.

The requirement that a distance indicator be testable against another
method is fundamental to the concept of the scientific method. 
Because we never know the ``right answer'' in the distance scale business,
we turn to intercomparisons between different methods assuming that
2 independent methods are very unlikely to repeatedly yield the
same wrong answer. If a method is not testable, it relies solely
on the validity of a model and scientists generally agree that models must
be tested. By inference, an untestable indicator is equivalent
to an untestable model. Fortunately, the PNLF can be tested against
numerous methods (see \cite{CJT93}, \cite{Jacoby95}, \cite{Feldmeier97}).

The second component of the last requirement is that multiple
investigators must be able to derive the same answer using the same
technique. This sounds simple enough, and again, is fundamental to the
scientific method.  Results from some methods, however, cannot be
reproduced at a later time should a question arise about their
validity.  The most obvious of these methods is supernovae. A second
observer cannot go back in time to observe a supernova in order to
check on the observational accuracy of a prior observer's measurements,
or to utilize a superior instrument. Thus, SN Ia fail to satisfy this
requirement.

The PNLF technique satisfies the prescription for a good distance indicator
on each count. Cross-testing with other methods (the last and most 
important criterion) shows that disagreements between the PNLF and other
reliable methods (\eg Cepheids) are smaller than 8\%. Thus, systematic
errors due to extinction, age, metallicity, or application of the method
are not accumulating beyond the 8\% level. In fact, when consideration
is made for the error contribution from the Cepheid distances, the PNLF
errors must be smaller than $\sim$5\%.

\begin{table}
\begin{center}
\begin{tabular}{llrrl}
{\bf Galaxy} & {\bf Type} & {\bf Nr.~PN} & {\bf (m-M)$_0$} & {\bf Reference} \\
     &    &    &               & \\
\multicolumn{5}{l}{\bf Local Group} \\
LMC &SBm &42 &$18.44 \pm 0.18$ & \cite{JWC} \\
SMC &Im &8 &$19.09 \pm 0.29$  & \cite{JWC} \\
185 &dE3p &4 &... &  \cite{Ciardullo89} \\
205 &S0/E5p &12 &$24.68 \pm 0.35$ & \cite{Ciardullo89} \\
221 &E2 &9 &$24.58 \pm 0.60$ & \cite{Ciardullo89} \\
224 &Sb &104 &$24.26 \pm 0.04$  & \cite{Ciardullo89} \\
     & &    &    &              \\
\multicolumn{5}{l}{\bf NGC 1023 Group} \\
891 &Sb &34 &$29.97 \pm 0.16$ & \cite{CJH91} \\
1023 &SB0 &97 &$29.97 \pm 0.14$ & \cite{CJH91} \\
     & &    &    &               \\
\multicolumn{5}{l}{\bf Fornax Cluster} \\
1316 &S0p &58 &$31.13 \pm 0.07$ & \cite{MCJ93} \\
1399 &E1 &37 &$31.17 \pm 0.08$  & \cite{MCJ93} \\
1404 &E2 &19 &$31.15 \pm 0.10$  & \cite{MCJ93} \\
     &    &    &               & \\
\multicolumn{5}{l}{\bf Leo I Group} \\
3377 &E6 &22 &$30.07 \pm 0.17$  & \cite{CJF89} \\ 
3379 &E0 &45 &$29.96 \pm 0.16$  & \cite{CJF89} \\
3384 &SB0 &43 &$30.03 \pm 0.16$ & \cite{CJF89} \\
3368 &Sab &25 &$29.91 \pm 0.15$ & \cite{Feldmeier97} \\
     &    &    &               & \\
\multicolumn{5}{l}{\bf Virgo Cluster} \\
4374 &E1 &37 &$30.98 \pm 0.18$  & \cite{JCF90} \\
4382 &S0 &59 &$30.79 \pm 0.17$  & \cite{JCF90} \\
4406 &S0/E3 &59 &$30.98 \pm 0.17$ & \cite{JCF90} \\
4472 &E1/S0 &26 &$30.71 \pm 0.19$ & \cite{JCF90} \\
4486 &E0 &201 &$30.73 \pm 0.19$  & this paper \\
4649 &S0 &16 &$30.76 \pm 0.19$  & \cite{JCF90} \\
     &    &    &               & \\
\multicolumn{5}{l}{\bf Coma I Group} \\
4278 &E1 &23 &$30.04 \pm 0.18$  & \cite{JCH96} \\
4494 &E1 &101 &$30.54 \pm 0.14$ & \cite{JCH96} \\
4565 &Sb &17 &$30.12 \pm 0.17$  & \cite{JCH96} \\
     &    &    &               & \\
\multicolumn{5}{l}{\bf NGC 5128 Group} \\
5102 &S0p &19 &$27.47 \pm 0.22$  & \cite{MCJ94} \\
5128 &S0p &224 &$27.73 \pm 0.04$ & \cite{Hui93} \\
5253 &Amorph &16 &$27.80 \pm 0.29$ & \cite{Phillips92} \\
     &    &    &               & \\
\multicolumn{5}{l}{\bf Other} \\
Bulge &Sbc &22	&$14.54 \pm 0.20$  & \cite{Pottasch90} \\
300  &Sc &10 &$26.78 \pm 0.40$  & \cite{Soffner96} \\
3031 &Sb &88 &$27.72 \pm 0.25$  & \cite{JCFB} \\
3109 &Sm &7 &$26.03 \pm 0.30$  & \cite{RM92} \\
3115 &S0 &52 &$30.11 \pm 0.20$  & \cite{CJT93} \\
4594 &Sa &204 &$29.76 \pm 0.13$ & \cite{Ford96} \\
5194/5 &Sbc/SB0 &38 &$29.56 \pm 0.15$  & \cite{Feldmeier97} \\
5457 &Sc &27 &$29.36 \pm 0.15$  & \cite{Feldmeier97} \\
\end{tabular}
\end{center}
\medskip
\caption{All distances must be increased by 0.06 mag to be placed on
the recent M31 Cepheid scale of Freedman \& Madore (1990).
A metallicity correction (Ciardullo \& Jacoby 1992) has been applied 
to NGC 5253 only, because it has SMC-like abundances.}
\end{table}

\section{Some New Insights}

If we accept that the PNLF method yields accurate distances, it seems
odd, at first, that the errors from population differences and extinction
aren't larger. A potentially dominant population effect is age, as  
discussed already. The key point is that intermediate age populations
all produce nearly identical central star masses. This follows from
the initial-to-final mass relation (\cite{Weidemann87}). That is,
for progenitor masses between 1 and 2 M$\subsun$ corresponding to ages
of about 1 to 10 Gyr, the central star mass will be in the narrow
range of $\sim 0.58\pm0.02$ M$\subsun$.  This narrow range is close
to that observed for white dwarfs (\cite{McMahon89}).

Another important effect arises in young ($<$0.5 Gyr) populations to
inhibit [OIII] luminous PN from forming. \cite{KJ91} showed, and
\cite{Dopita96} confirmed, that for young progenitors producing central
stars more massive than 0.65 M$\subsun$, the surface abundances are
strongly altered such that nitrogen is greatly enhanced.  The added
nitrogen competes with oxygen in cooling the nebula, to the detriment
of the [OIII] luminosity. Thus, PN deriving from young, massive
progenitors fail to populate the high luminosity end of the PNLF and
the effect of a young population on the PNLF is lost.

Similarly, metallicity seems like it ought to play a large role. A
competition exists, though, between the efficiency of the nebula to
radiate in [OIII] and the luminosity input from the central star. 
Higher metallicity values enhance the nebula's ability to
radiate at $\lambda5007$.  The central star, however, is predicted to
have a lower mass and luminosity as a consequence of experiencing
higher mass loss prior to leaving the AGB. The reduced heating 
compensates to first order for the enhanced radiative efficiency 
as metallicity increases (\cite{Dopita92}).

\section{Summary of PNLF Distances}

Table 1 summarizes the available PNLF distances.
Distances are referenced to a zero-point based on M31 having
a distance of 710 kpc and a reddening of E(B--V)$ = 0.11$. This is
the baseline zero-point used in all our papers. To place
these distances on the recent scale where the distance to M31
is 770 kpc (\cite{Freedman90}) and E(B--V)$ = 0.08$, all distances
should be increased by 0.06 mag (3\%).

Ten galaxies of the 34 in this list are obvious spirals. 
Six more have a significant late-type component for which Cepheid
distances either exist already or could be determined. 

\section{Spirals}

Extending the PNLF technique to spiral galaxies requires extra care due
to 3 factors. First, potential confusion exists between PN and HII
regions. Second, spiral arms contain obvious dust lanes that could
reduce observed luminosities.  Third, a young population of stars
must exist in spiral arms.  Nevertheless, the advantage of working in
spirals is tremendous because their Cepheids provide the most accepted
reference distances to test the PNLF method for systematic errors.
Also, spirals are where zero-point calibrations are most believed.

To identify PN in spiral galaxies, we complement the $\lambda5007$
on-band/off-band imaging technique with on-band/R-band images at
H$\alpha$.  We define the following criteria for PN candidates in spirals.
A PN candidate must:

\medskip
\begin{enumerate}
\renewcommand{\theenumi}{\arabic{enumi}}
\item have a stellar PSF
\item be on the [OIII] image and absent on a continuum image
\item be absent in an R-band image
\item be absent or extremely weak in an H$\alpha$ image
\item not be in a spiral arm
\end{enumerate}
\medskip

\noindent
These criteria have been applied recently to observations of M51 (NGC 5194),
M96 (NGC 3368), and M101 (NGC 5457) by \cite{Feldmeier97}.  M96 and M101 
distances exist already from HST Cepheid surveys.

\begin{figure}
\epsfysize 432pt \epsfbox{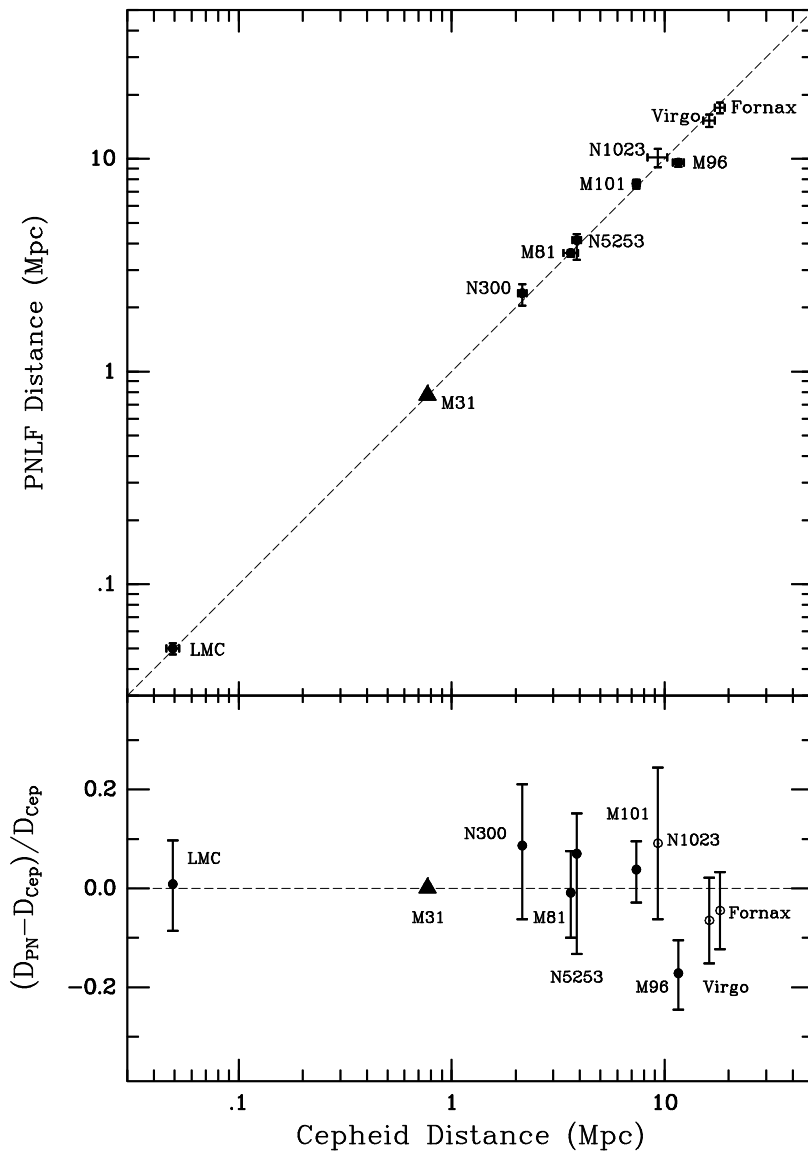}
\caption{A comparison between PNLF and Cepheid distances. Solid circles
represent direct galaxy comparisons; the solid triangle is the calibrator
galaxy, M31, which, by definition, falls exactly on the dashed 1:1 line.
The points labeled N1023, Virgo, and Fornax represent comparisons
between elliptical (PNLF) and spiral (Cepheid) galaxies within the
same cluster. The lower panel  clarifies the level of disagreement
by plotting the relative differences in distances. Only M96 deviates
by more than 1$\sigma$; at this time, it is unclear if the disagreement 
is significant, and if so, whether the PNLF or Cepheid distance is
discrepant.}
\end{figure}

Figure 1 illustrates the excellent agreement between the PNLF (adjusted to
the modern M31 distance and extinction) and Cepheid distance scales. 
The PNLF distances are based solely on M31 as the zero-point calibrator,
yet the mean offset for the additional 6 Cepheid galaxies differs by 
only 1\%. The random scatter of 8\% about the ridge line is consistent
with, or smaller than, the combined errors of the Cepheid and PNLF distances.
Note that some uncertainties are common to both methods (e.g., foreground
extinction) which slightly reduce the apparent uncertainty in the
combined distance error.

Figure 1 also includes comparisons between PNLF distances to elliptical
galaxies and Cepheid distances for different galaxies in the same
group.  These indirect comparison points provide further support, but
do not confirm the PNLF distances to the Virgo, Fornax, Leo, and NGC
1023 groups.

The lack of any evidence for systematic errors, either among the direct
or indirect comparisons, strongly implies that any residual systematic
errors for zero-points, population effects (age and metallicity),
internal extinction, and methodology (adopted PNLF shape, sizes of PNLF
samples) must be smaller than $\sim$8\%. Alternatively, one can insist
that a conspiracy exists among these parameters such that both Cepheid
and PNLF distances have errors that correlate.

\section{The Spiral--Elliptical Connection}

To a limited degree, the 3 indirect distance comparisons shown in
Figure 1 illustrate that the PNLF distances to ellipticals are on
the same system as the spiral distances. In addition, tests relying
purely on PNLF distances have been described for the NGC 1023 group
(\cite{CJH91}), the Coma I group (Jacoby, Ciardullo, and Harris 1996), 
and the Leo group (\cite{Feldmeier97}). Figure 2 summarizes the 4 PNLF
and 2 Cepheid distances to the Leo group.

\begin{figure}
\epsfysize 216pt \epsfbox{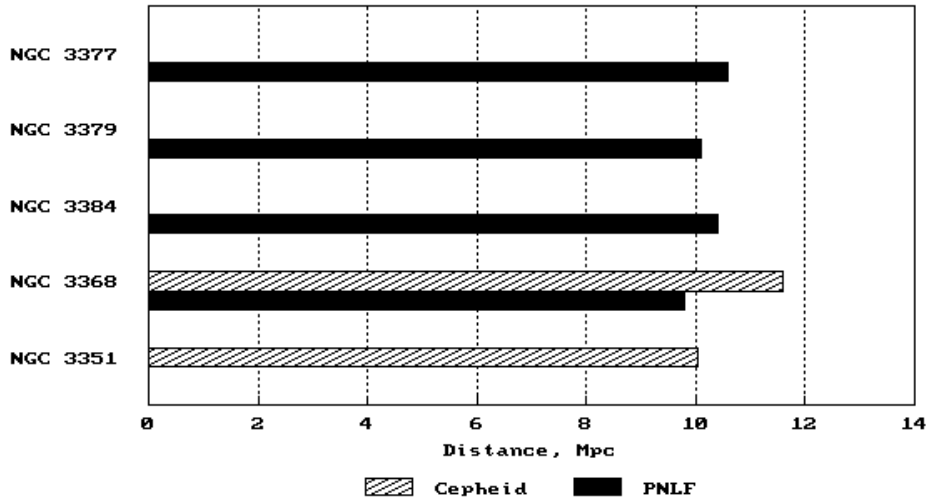}
\caption{A comparison between PNLF and Cepheid distances in the Leo Group. 
NGC 3368 (M96) has been measured using both the PNLF (Feldmeier \etal (1997))
and with Cepheids (Tanvir \etal (1995)). In addition, group member NGC 3351 
has been measured with Cepheids (Graham \etal (1996)).}
\end{figure}

Tests to different galaxies cannot be conclusive, since the two
galaxies are not guaranteed to be at exactly the same distance.
Nevertheless, the deviations in distance (spiral minus elliptical) for
the 3 purely PNLF tests are as follows. NGC 1023 group: $0.00\pm0.21$
mag (NGC 1023); Coma I group: $+0.08\pm0.15$ mag (NGC 4278) and
$-0.42\pm0.12$ mag (NGC 4494); Leo group: $-0.12\pm0.24$ mag (NGC
3377), $-0.11\pm0.24$ mag (NGC 3379), and $-0.18\pm0.24$ (NGC 3384).
At face value, assuming that in each of these 3 cases the spiral and
associated ellipticals are at exactly the same distance, the average
offset is $-0.16$ mag for the 3 cases (0.00 for NGC 1023, $-0.34$ for Coma
I, and $-0.13$ for Leo) in the sense that our elliptical distances are
too large.

This conclusion is too simplistic because we have not addressed
the assumption that the spiral galaxy is always at the same distance
as the comparison galaxy. For the Coma I group, for example, the GCLF
(\cite{Fleming95}) and the SBF (\cite{SP94}) methods concur that NGC 4494 
is beyond NGC 4565, and curiously, the deviance for this galaxy is the
largest we see. If removed from the sample, the spiral--elliptical offset
is reduced to $-0.02$ mag, a level that is too small to consider reliable
within the assumptions. Until this problem can be addressed more
thoroughly, we apply no correction to our elliptical galaxy
distances.

\section{Details of New M87 PNLF Studies}

Bottinelli \etal (1991) and Tammann (1992) argued that PNLF distances to
Virgo ellipticals were underestimated because the luminosity depth of the PN
surveys was not adequate to sample beyond the brightest (0.5 mag) edge
of the PNLF.  Since that edge is nearly linear in the logarithmic PNLF,
the method becomes insensitive to distance modulus. In addition, those
authors challenged the PNLF distances on the basis that a shallow
survey of a large galaxy will suffer from a sample size bias. The sense
of this argument is that $N$ objects are more likely to be drawn from
the low probability bright tail of the large elliptical galaxy PN
sample than are $N$ objects from the smaller sample in M31's bulge.

While it is true that PN surveys must extend deep enough to sense the
curvature of the PNLF reliably with statistical methods, the required
depth is only 0.8 mag.  With the exception of NGC 4649 which
was observed under poor conditions, Jacoby, Ciardullo, \& Ford (1990)
estimated the depth of their surveys for 6 Virgo galaxies to be $\sim
1.0$ mag. Thus, it seemed unlikely that a serious systematic error was
contaminating those distances.

An independent assessment of the likelihood of a serious systematic
error is provided by recent Cepheid distances to Virgo galaxies.
The average PNLF distance to Virgo, based on 6
galaxies, is 15.3 Mpc (using the modern M31 distance and extinction for
the zero-point). This result agrees very well with the
\cite{Ferrarese96} distance of 15.8 Mpc to M100 based on HST Cepheids,
and the \cite{Pierce94} distance of 14.7 Mpc to NGC 4571 based on CFHT
Cepheids. In addition, \cite{Sandage96} reports HST Cepheid distances
to three near-Virgo galaxies: NGC 4496 at 16.6 Mpc, NGC 4536 at 16.6
Mpc, and NGC 4639 at 25.1 Mpc. Thus, four galaxies are reported in the
range 14.7 to 16.6 Mpc, and these are very comparable to the PNLF range
of distances (14.3 to 16.2 Mpc). One galaxy, though, is behind
all of these. It is unclear which, if any, of these spirals
represents the distances to the ellipticals,
but it is evident that most of the spirals (four out of five) have
distances that support the PNLF distances.

A direct resolution of the challenges to the PNLF distances lies in a
short observing project. Deep PN observations in M87 can push well into
the plateau region of the PNLF. Data were obtained in April 1995 with
the KPNO 4-m telescope to examine the claim that the earlier Virgo
data were not deep enough.  A total of 7 hours of on-line integration
were devoted to detecting fainter PN. This survey also extends to large
radial (10 arcmin) distances from M87's nucleus. The survey results are
described below.

\subsection[]{A Deep PNLF Distance Contradicts the Challenges}

Figure 3 shows the new PNLF for M87. A total of 320 PN were identified,
but many are fainter than the completeness limit. A total of 201 PN
are in the complete sample which extends $\sim 1.2$ mag down the PNLF.

\begin{figure}
\epsfysize 216pt \epsfbox{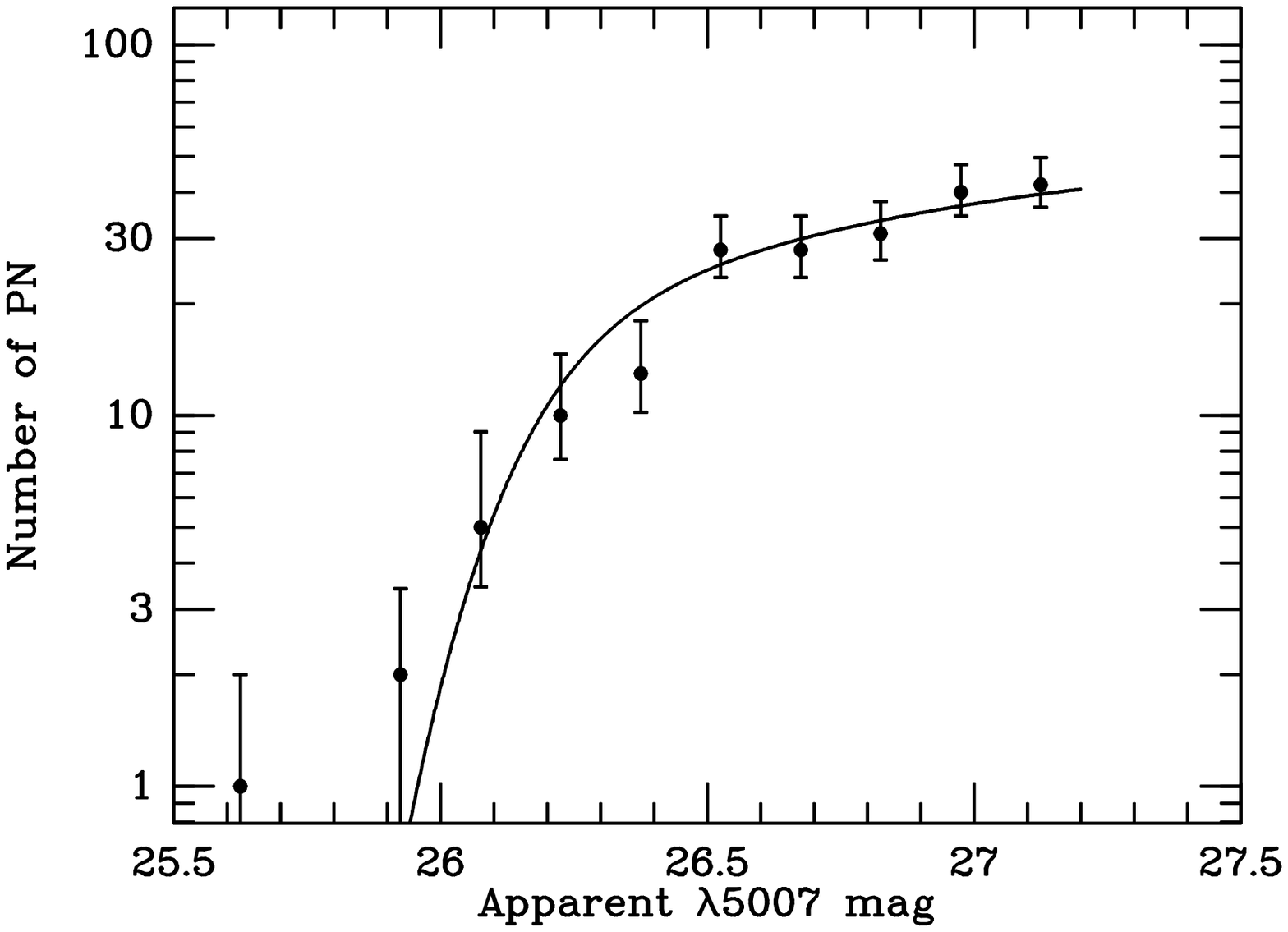}
\caption{The 1995 PNLF for M87 based on 7 hours at the KPNO 4-m.
This PNLF extends 1.2 mag down the PNLF and clearly reaches beyond
the linear portion of the bright edge ($25.9 < m_{5007} < 26.3$)
of the PNLF. The scaled and shifted
PNLF from M31 is superposed to illustrate the agreement with the reference
galaxy's PNLF and demonstrates that the M87 PNLF is not a power law
over this regime. Note the single very luminous
object at $m_{5007}=25.6$ which was found first in the 1990 survey.
See Jacoby \etal (1996) for a discussion of what this luminous object may be.}
\end{figure}

Figure 4 shows a curious effect, though, which has not been fully
evaluated at the time of this conference. That is, when the sample of
PN is divided in half such that the PN drawn from M87's halo are
separated, the PNLF for the inner half (out to 4 arcmin in radius) are
systematically fainter by 0.3 mag than the halo sample. I return to
this point below, but note here that the more reliable distance is
derived from the inner sample because it is less likely to be contaminated
by intracluster PN. That distance, $14.4\pm1.3$ Mpc (on the modern
M31 distance scale), is nearly identical to the \cite{JCF90} value of
$14.9\pm1.2$ Mpc.

\begin{figure}
\epsfysize 216pt \epsfbox{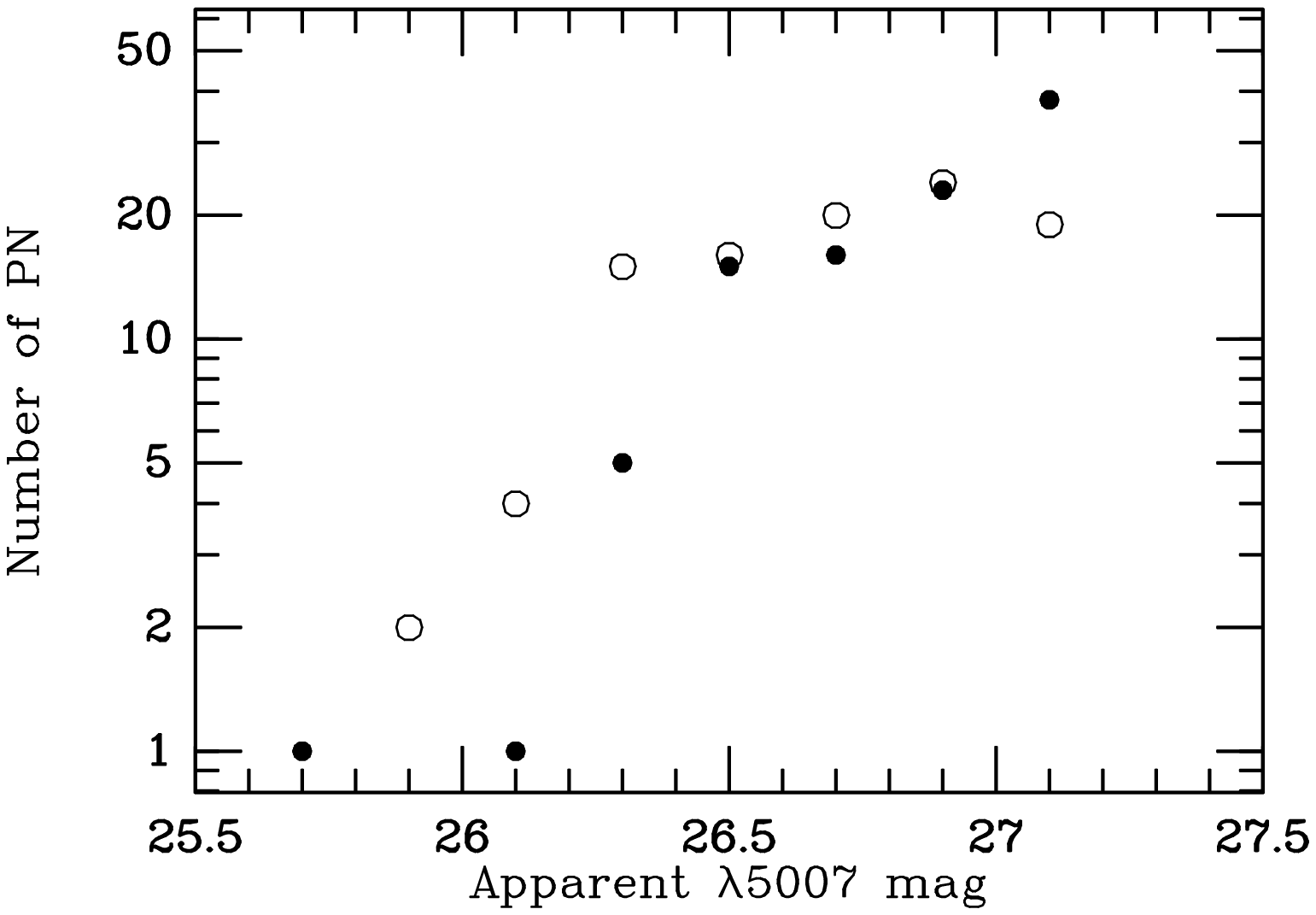} 
\caption{The 1995 PNLF for M87 where open circles represent those PN
found in the inner 4 arcmin and triangles show those PN found beyond 4 arcmin
and out to 10 arcmin.}
\end{figure} 

Thus, a deeper PNLF argues {\it against} the contentions of \cite{Bottinelli91} 
and \cite{Tammann92} that the PNLF distance to Virgo is underestimated
as a consequence of inadequate survey depth.

\subsection[]{Sample Size Effects are Small}

With the new larger sample of PN, it is possible to investigate the
effects that different sample sizes have on the final PNLF distance.
Subsets of PN were drawn randomly from the sample of 201 PN.  Distances
for these subsamples of 20, 30, 40, 50, and 100 PN were derived
following our standard procedures and compared to the distance based on
201 PN. Figure 5 shows the magnitude of the effect of sample size
differences. 

\begin{figure}
\epsfysize 216pt \epsfbox{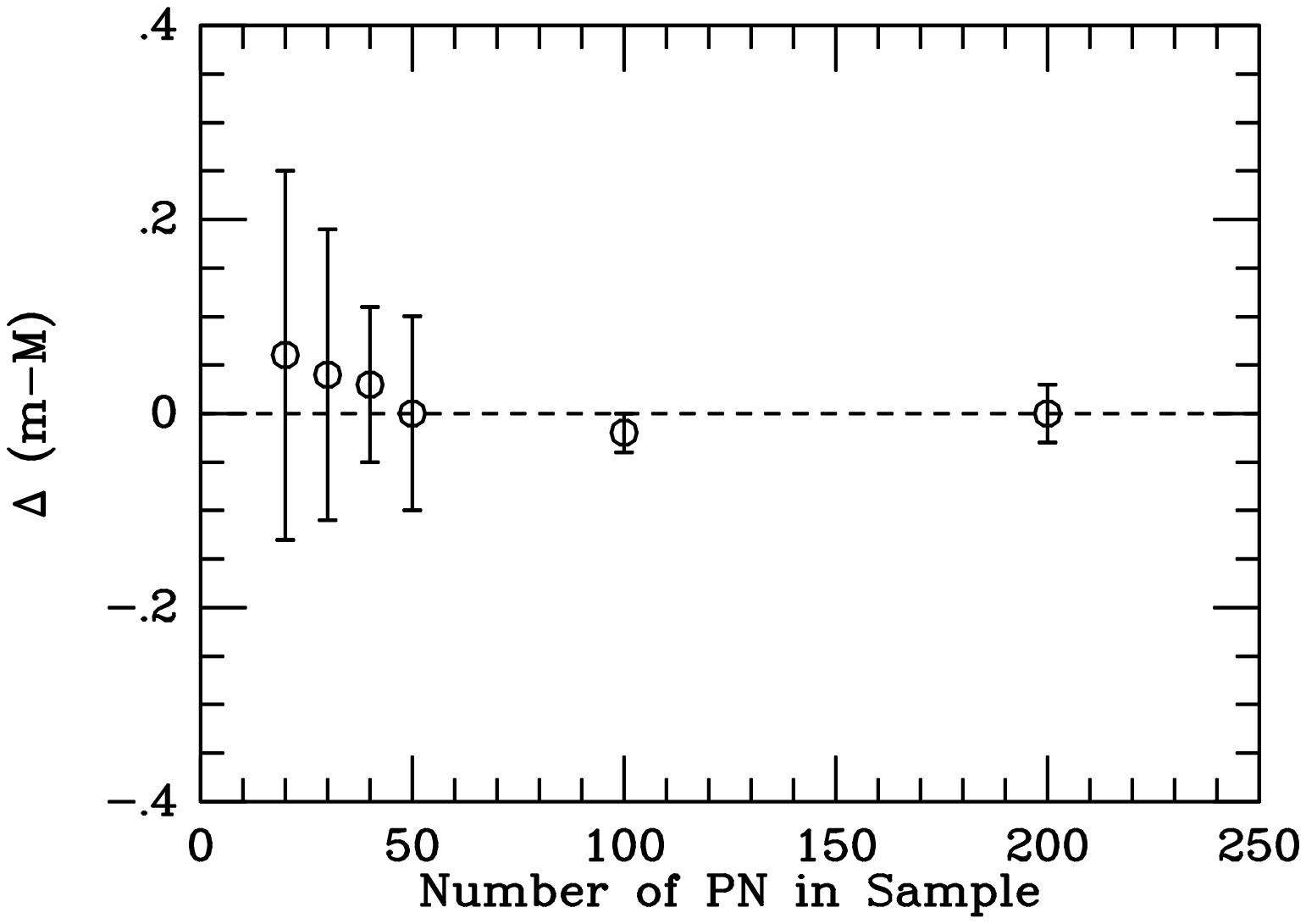}
\caption{Results of a Monte Carlo experiment to estimate the effect
of deriving distances with different sample sizes. The error bars
represent the scatter in the multiple attempts to derive distances with
a given number of PN in the sample.}
\end{figure}

In the worst case, for a sample of 20 PN, there is a slight tendency to
overestimate the distance to a galaxy by up to 3\%. Again, this
contradicts the challenges of \cite{Bottinelli91} and \cite{Tammann92}
who claim that our distances would be {\it underestimated}. The
reason that the effect is small is that the statistical process described
by \cite{Ciardullo89} is cognizant of the sample size and adjusts the
derived distances to the most likely one for a given sample. That is,
a statistical correction for sample size has always been applied to the
PNLF results.

\subsection[]{Bright PN in M87's Halo}

As noted above, M87's halo PNLF is $\sim 0.3$ mag brighter than
the central PNLF. We can all agree that the halo of M87
is {\it not} 15\% closer than its core! Thus, something must be artificially
enhancing the brightness of the PN in the outer regions. Since we have
not seen this effect in the two other samples that permit a similar
radial test (Cen A, \cite{Hui93}; NGC 4494, \cite{JCH96}), we 
consider what could cause such an effect here. Five possibilities
come to mind:

\medskip
\begin{enumerate}
\renewcommand{\theenumi}{\arabic{enumi}}
\item Metallicity decrease in halo
\item Age decrease in halo
\item Dust near center
\item Instrumental effect
\item Intracluster PN contamination
\end{enumerate}
\medskip

The color gradient in ellipticals is such that the outer halos are
bluer than the inner regions. The first 3 possibilities above
have been suggested as possible causes of gradients. The first,
a metallicity decrease, reduces the luminosity of PN, in contradiction
to the observed effect. The second, the presence of a young population
can enhance the PNLF luminosity if the ages of the halo stars are $<$0.5 Gyrs,
provided the central population is $>$3 Gyrs. We can
neither dismiss nor confirm this possibility.

The third option, dust in the central regions, has been suggested by
Wise \& Silva (1996), \cite{GD95}, and \cite{WTC92} to explain the
color gradients.  To explain the PNLF enhancement, central reddening
of E(B--V)$\sim$0.06 is needed. Again, this is plausible.

Although the effect has not been seen before, a brightness enhancement
in the halo could be caused by instrumental effects. Thus far, 
flat-fielding errors and filter transmission variations have been dismissed.

The fifth option initially seems highly speculative. The key point is
that a large population of intracluster stars exists and can produce PN
having a range of distances representing the full depth extent of the
Virgo cluster. Thus, some PN may be foreground to M87
and appear brighter than M87's own PN, while other PN
will be in the background and be lost in the faint end
of the PNLF.  Since the number of intracluster PN found is proportional
to the area surveyed, there is a survey bias against finding foreground
PN in the smaller central region, while simultaneously, there is a bias
in favor of finding true M87 PN at the center where the stellar density
is high.

What is the likelihood of finding intracluster PN? \cite{Arnabaldi}
found that 3 of 19 PN in the \cite{JCF90} NGC 4406 sample are intracluster.
Since that sample is biased against intracluster PN due to velocity
rejection in the survey filter (NGC 4406 was sampled at its systemic
velocity of $-220$ km/s, or about 1500 km/s from the Virgo systemic velocity),
there are likely to be many intracluster PN. \cite{Arnabaldi} discuss
the intracluster population in detail. To zeroth order, the number density
of intracluster PN is not a problem.

A definitive source of the enhanced halo PNLF in M87 is not possible
at this time. Kinematics can be used to investigate the likelihood
that intracluster PN are contaminating the halo sample.
The other likely causes, extinction and very young populations, seem
less secure at this time because we don't know that they exist while
we do know that intracluster PN do exist.

\section[]{Can M87 Really Be At 22 Mpc?}

Sandage and Tammann (e.g., \cite{Sandage90}) have long argued that for
values of H$_0 < $55, the Virgo Cluster must be at a distance of about
22 Mpc since the cosmic velocity of Virgo is $\sim 1170$ km/s.
Additionally, M87 is usually claimed to be representative of the Virgo
distance since its recessional velocity is close to the Virgo average,
it is projected near the center of the cluster, and it appears to be
at the bottom of the potential well as evidenced by
an x-ray cloud typical of x-ray clusters.

A reasonable question to ask is ``what must one give up to
derive a PNLF distance of 22 Mpc for M87?'' The simple answer is
that the constancy of the maximum PN luminosity must be abandoned. And,
it must be abandoned in a big way -- by $\sim$0.9 mag.

The astrophysical question becomes ``can the PNLF be enhanced by 0.9 mag,
and if so, what are the observable manifestations?'' The latter turn out to
be severe and obvious. Models of the PNLF by \cite{Jacoby89}, \cite{Mendez93},
and \cite{Stanghellini95} all show that a shift in the PNLF as large
as 0.9 mag is almost impossible because there is inadequate UV flux
being radiated by any reasonable collection of central stars.  A 0.9
mag enhancement demands that the central stars be very massive,
originating from 4--5 M$\subsun$ progenitors. A population this young
($<100$ Myrs) should be evident.

The presence of a very young population in ellipticals is not
unreasonable; many ellipticals exhibit evidence for gas and dust
(\cite{Goudfrooij}). Usually, though, the mass fraction of the
elliptical involved in the young population is tiny. To enhance the
PNLF by 0.9 mag, though, requires that $\sim$40\% of the total
luminosity of the galaxy comes from the young population.  (The 40\%
value derives from the PN production rate for bright PN in M87.  This
rate is 40\% of the maximum value that is attainable under the
assumption that all stars produce PN. Since some stars may not produce
PN, the actual fraction of luminosity from young stars could be higher.)

A trivial observational test for a young population is galaxy color.
A young population ($<$100 Myr) has a color (B--V) $<$ 0.0 while an old
population ($>$11 Gyr) has a color (B--V) $\sim$ 1.0. Thus a mix where
40\% of the luminosity derives from a young population has a color
(B--V) $\sim$ 0.6. This color is the direct consequence of insisting that
M87 be at 22 Mpc, combined with stellar evolution predictions 
(e.g., \cite{Ciotti}); \cite{Vass})
for the initial-to-final mass relation that was devised to explain
the observed relation (\cite{WK83}). 

Comparing the predicted color of 0.6 to the observed (B--V) $\sim$ 1.0
(which depends slightly on radial position), it is evident that a sizable
young population is untenable. Thus, we are forced to accept that either
M87 is much closer than 22 Mpc in order to alleviate the pressure to
enhance the PNLF luminosity, or that the initial-to-final mass relation
predicted by stellar evolution models and observed in young Galactic clusters
is seriously flawed in such a way that old stars can produce massive
central stars. Since there is no evidence for the latter, the more
likely conclusion is that the existence of PN in M87 at the observed
apparent magnitudes demands a distance much smaller than 22 Mpc.
(It is possible to push the distance as large as 17 Mpc
before the implied population color becomes inconsistently blue.)

\section{Some Important Distances}

Based on the comparisons between PNLF and Cepheid distances, between
PNLF and SBF (\cite{CJT93}) distances, and between PNLF spiral and PNLF
elliptical distances within a cluster, there is repeated evidence that
PNLF distances are accurate to $\sim$8\%.  Similar claims are being
made for SN Ia.  Yet, PNLF distances suggest values for H$_0$ near 75
km/s/Mpc while SN Ia distances are between 55 and 67 km/s/Mpc
(\cite{Sandage96}; Riess \etal (1996); Hamuy \etal (1996)) and both
methods are based on Cepheids for calibration. While an H$_0$ of 67 is
within the combined error budget of the methods, an H$_0$ of 55
stresses the PNLF method (and others) severely.

The PNLF and SN Ia distance comparison shown in Table 2 for galaxies in
common in the Virgo and Fornax clusters summarizes the issue.  For
Virgo, these include a single SN Ia in each of NGC 4374, NGC 4382, and
NGC 4486, although none was observed very well.  Although SN1991bg in NGC 4374
was observed very well, we reject it for being anomalously
underluminous (\cite{Phillips93}; \cite{Leib93}).  For Fornax, 2 SN Ia
occurred in NGC 1316, and both were reasonably well observed
(\cite{Hamuy91}).  Distances are given as the average of the 3 galaxies
in Virgo.

The SN Ia zero-point is set using the well observed SN Ia SN1981B and
SN1990N. No others (SN1895B, SN1937C, SN1960F, SN1972E) were adequately
observed for their peak magnitudes and decline rates to be determined
at the same level of reliability as these 2 primary calibrators. From
\cite{Sandage96}, we have M$_{\rm B}^{max} = -19.3\pm0.08$ for SN1981B
and SN1990N. The apparent peak magnitudes for the Virgo and Fornax SN
Ia are taken from \cite{Leib91}.

\begin{table}
\begin{center}
\begin{tabular}{lccc}
Cluster & PNLF Distance & SN Ia Distance & Discrepancy     \\
        &   Mpc         &     Mpc        &   in $\sigma$'s \\
        &               &                &                 \\
Virgo   & $15.1\pm0.9$  &  $17.4\pm2.6$  &    0.8$\sigma$  \\
Fornax  & $17.7\pm0.5$  &  $22.9\pm1.5$  &    3.3$\sigma$  \\
\end{tabular}
\end{center}
\caption{Comparison Between PNLF and SN Ia Distances to Virgo and Fornax}
\end{table}

Evidently, the discrepancy in Virgo is not significant while the
discrepancy in Fornax is significant. Thus, either the PNLF distance to
Fornax is incorrect (although 2 other galaxies yield the same distance
and the Cepheid distance is nearly identical), or the 2 well observed SN
Ia in Fornax are underluminous by 0.5 mag.  There is evidence that the
Fornax SN Ia are, in fact, somewhat fast declining ones, but this can
only explain about 0.15 mag of the discrepancy. If the decline rate is
considered, using the \cite{Hamuy96} slope of 0.78, then the
discrepancy is reduced to a 2.2$\sigma$ event, which still is
marginally significant.  Because the discrepancies between the Fornax
SN Ia distance and Fornax distances from the HST Cepheids, SBF, and
PNLF are all very similar, the SN Ia Fornax inconsistency 
cannot be solved by appealing solely to errors in the PNLF method.
For now, this issue remains open.

\section{Conclusions} 

No distance indicator is perfect. Confidence in the results, though,
is built by extensive testing against other distance indicators,
and through the development of the astrophysical theory supporting
the indicator's use.

The PNLF technique has been well tested against Cepheids, SBF, and against
itself in clusters. Constancy of the bright end of the PNLF has been 
easy to reproduce theoretically if the ages of the stars producing PN are 
within a plausibly wide range. Thus, PNLF distances
appear reliable at this time. The main points of this paper are:

\medskip
\begin{enumerate}
\renewcommand{\theenumi}{\arabic{enumi}}
\item PNLF distances to spirals are consistent with Cepheid distances.
\item There are now 7 PNLF calibrators; their dispersion is 8\%.
\item Good agreement with Cepheid and SBF distances must mean that
systematic errors (due to age, metallicity, extinction, PN sample sizes,
PNLF methodology) are under control to the limits of the deviations
of these methods (typically, 8\%).
\item The M87 deep survey demonstrates that the bright end of the PNLF
is not a power law as some have maintained and that PNLF distances are
not sensitive to the number of PN in the sample.
\item Both the Virgo and Fornax distances agree with the Cepheid-based
distances, but the Fornax SN Ia distance is $\sim$25\% larger. The Virgo
SN Ia distance is not discrepant, within the errors.
\end{enumerate}
\medskip

In terms of the future application of the PNLF, it is somewhat costly
in telescope time (exposure time is proportional to distance$^4$).
So, while the method yields results as reliable as the Cepheids or SBF,
its application may be reserved for galaxies that present problems
for other methods. For example, it can be used to measure distances
to SN Ia hosts such as Sa/S0 galaxies where Cepheids won't be found easily
and where galaxy structure compromises the SBF.

\medskip
\medskip

\noindent
{\bf Acknowledgements}
\begin{acknowledgments}
This paper derives from other papers in press. The discussions
on spirals was based on information provided by John Feldmeier; preliminary
results on the deep M87 survey were provided by Robin Ciardullo.
I am grateful to John Graham who gave me his HST Cepheid distance to 
NGC 3351 prior to publication.
\end{acknowledgments}

\clearpage



\begin{thebibliography}{} 

\bibitem[Arnabaldi \etal (1996)]{Arnabaldi}
     {\sc Arnabaldi, M., Freeman, K. C., M\'endez, R. H., Capaccioli,
     M., Ciardullo, R., Ford, H., Gerhard, O., Hui, X., Jacoby, G. H.,
     Kudritzki, R. P., \& Quinn, P. J.} 1996
     {\it Ap.J.}, submitted.

\bibitem[Bottinelli \etal (1991)]{Bottinelli91}
     {\sc Bottinelli, L., Gouguenheim, L., Paturel, G., \& Teerikorpi, P.} 1991
     {\it Astr.Ap.} {\bf 252}, 550.

\bibitem[Ciardullo \& Jacoby (1992)]{CJ92}
     {\sc Ciardullo, R. \& Jacoby, G. H.} 1992 
     {\it Ap.J.} {\bf 388}, 268--271.

\bibitem[Ciardullo, Jacoby, \& Ford (1989)]{CJF89}
     {\sc Ciardullo, R., Jacoby, G. H., \& Ford, H. C.} 1989
     {\it Ap.J.} {\bf 344}, 715--725.
 
\bibitem[Ciardullo \etal (1989)]{Ciardullo89}
     {\sc Ciardullo, R., Jacoby, G. H., Ford, H. C., \& Neill, J. D.} 1989 
     {\it Ap.J.} {\bf 339}, 53--69.

\bibitem[Ciardullo, Jacoby, \& Harris (1991)]{CJH91}
     {\sc Ciardullo, R., Jacoby, G. H., \& Harris, W. E.} 1991 
     {\it Ap.J.} {\bf 383}, 487--497.

\bibitem[Ciardullo, Jacoby, \& Tonry (1993)]{CJT93}
     {\sc Ciardullo, R., Jacoby, G. H., \& Tonry, J. L.} 1993 
     {\it Ap.J.} {\bf 419}, 479.

\bibitem[Ciotti \etal (1991)]{Ciotti}
     {\sc Ciotti, L., D'Ercole, A., Pellegrini, S., \& Renzini, A.} 1991
     {\it Ap.J.} {\bf 376}, 380.

\bibitem[Dopita \etal (1992)]{Dopita92} 
     {\sc Dopita, M. A., Jacoby, G. H., \& Vassiliadis, E.} 1992
     {\it Ap.J.} 27--38.

\bibitem[Dopita \etal (1996)]{Dopita96}
     {\sc Dopita, M. A., Vassiliadis, E., Wood, P. R., Meatheringham, S. J.,
     Harrington, J. P., Bohlin, R. C., Ford, H. C., Stecher, T. P., \& Maran,
     S. P.} 1996
     {\it Ap.J.}, in press.

\bibitem[Feldmeier, Ciardullo, \& Jacoby (1997)]{Feldmeier97}
     {\sc Feldmeier, J., Ciardullo, R., \& Jacoby, G. H.} 1997
     {\it Ap.J.} submitted.

\bibitem[Ferrarese \etal (1996)]{Ferrarese96}
     {\sc Ferrarese, L., \& others} 1996
     {\it Ap.J.} {\bf 464}, 568.

\bibitem[Fleming \etal (1995)]{Fleming95}
     {\sc Fleming, D. B., Harris, W. E., Pritchet, C. J., \& Hanes, D. A.} 1995
     {\it AJ} {\bf 109}, 1044.

\bibitem[Ford \etal (1996)]{Ford96}
     {\sc Ford, H. C., Hui, X., Ciardullo, R., Jacoby, G. H., \& 
     Freeman, K. C.} 1996
     {\it Ap.J.} {\bf 458}, 455.

\bibitem[Freedman \& Madore (1990)]{Freedman90}
     {\sc Freedman, W. L., \& Madore, B. F.} 1990
     {\it Ap.J.} {\bf 365}, 186.

\bibitem[Goudfrooij (1995)]{Goudfrooij}
     {\sc Goudfrooij, P.} 1995
     {\it P.A.S.P.} {\bf 107}, 502.

\bibitem[Goudfrooij \& De Jong (1995)]{GD95}
     {\sc Goudfrooij, P., \& De Jong, T.} 1995
     {\it Astr.Ap.} {\bf 298}, 784.

\bibitem[Graham \etal 1996]{Graham96}
     {\sc Graham, J. A., Phelps, R. L., Freedman, W. L., Saha, A.,
     Stetson, P. B., Madore, B. F., Silbermann, N. A., Sakai, S.,
     Kennicutt, R. C., Harding, P., Turner, A., Mould, J. R., Ferrarese,
     L., Ford, H. C., Hoessel, J. G., Han, M., Huchra, J. P., Hughes, S.
     M., Illingworth, G. D., \& Kelson, D. D.} 1996
     {\it Ap.J.}, submitted.

\bibitem[Hamuy \etal (1991)]{Hamuy91}
     {\sc Hamuy, M., Phillips, M. M., Maza, J., Wischnjewsky, M., 
     Uomoto, A., Landolt, A. U., \& Khatwani, R.} 1991
     {\it AJ} {\bf 102}, 208.

\bibitem[Hamuy \etal (1996)]{Hamuy96}
     {\sc Hamuy, M., Phillips, M. M., Schommer, R. A., Suntzeff, N. B.,
     Maza, J., \& Avil\'es, R.} 1996 
     {\it AJ} submitted.

\bibitem[Han \etal (1994)]{Han94} 
     {\sc Han, Z., Podsiadlowski, P., \& Eggleton, P. P.} 1994
     {\it M.N.R.A.S.} {\bf 270}, 121

\bibitem[Hui \etal 1993]{Hui93}
     {\sc Hui, X., Ford, H. C., Ciardullo, R., \& Jacoby, G. H.} 1993
     {\it Ap.J.} {\bf 414} 463

\bibitem[Jacoby (1989)]{Jacoby89} 
     {\sc Jacoby, G. H.} 1989
     {\it Ap.J.} {\bf 339}, 39.

\bibitem[Jacoby (1995)]{Jacoby95} 
     {\sc Jacoby, G. H.} 1995
     in {\it Science With the VLT} eds. J. R. Walsh \& I. J. Danziger, 267.

\bibitem[Jacoby, Ciardullo, \& Ford (1990)]{JCF90}
     {\sc Jacoby, G. H., Ciardullo, R., \& Ford, H. C.} 1990
     {\it Ap.J.} {\bf 356}, 332--349.

\bibitem[Jacoby \etal (1989)]{JCFB}
     {\sc Jacoby, G. H., Ciardullo, R., Ford, H. C., Booth, J.} 1989
     {\it Ap.J.} {\bf 344}, 70.

\bibitem[Jacoby, Ciardullo, \& Harris (1996)]{JCH96}
     {\sc Jacoby, G. H., Ciardullo, R., \& Harris, W. E.} 1996
     {\it Ap.J.} {\bf 462}, 1--12.

\bibitem[Jacoby, Walker, \& Ciardullo (1990)]{JWC}
     {\sc Jacoby, G. H., Walker, A. R., \& Ciardullo, R.} 1990
     {\it Ap.J.} {\bf 365}, 471--477

\bibitem[Jacoby \etal (1992)]{Jacoby92} 
     {\sc Jacoby, G. H., Branch, D., Ciardullo, R., Davies, R. L., Harris, 
      W. E., Pierce, M. J., Pritchet, C. J., Tonry, J. L., \& Welch, D. L.} 
     1992 
     {\it P.A.S.P.} {\bf 104}, 599--662.

\bibitem[Kaler \& Jacoby (1991)]{KJ91} 
     {\sc Kaler, J. B. \& Jacoby, G. H.} 1991
     {\it Ap.J.} {\bf 382}, 134.

\bibitem[Leibundgut \etal (1991)]{Leib91}
     {\sc Leibundgut, B., Tammann, G. A., Cadonau, R., \& Cerrito, D.} 1991
     {\it Astr.Ap.Sup.} {\bf 89}, 537--579.

\bibitem[Leibundgut \etal (1993)]{Leib93}
     {\sc Leibundgut, B. \& others} 1993
     {\it AJ} {\bf 105}, 301.

\bibitem[McMahon (1989)]{McMahon89}
     {\sc Mcmahon, R. K.} 1989
     {\it Ap.J.} {\bf 336}, 409.

\bibitem[McMillan, Ciardullo, \& Jacoby (1993)]{MCJ93}
     {\sc McMillan, R., Ciardullo, R., \& Jacoby, G. H.} 1993
     {\it Ap.J.} {\bf 416}, 62--73.

\bibitem[McMillan, Ciardullo, \& Jacoby (1994)]{MCJ94}
     {\sc McMillan, R., Ciardullo, R., \& Jacoby, G. H.} 1994
     {\it AJ} {\bf 108}, 1610--1618.

\bibitem[M\'endez \etal (1993)]{Mendez93} 
     {\sc M\'endez, R. H., Kudritzki, R. P., Ciardullo, R. P., \& Jacoby, G. H.}
     1993 
     {\it Astr. Ap.} {\bf 275}, 534--548.

\bibitem[Phillips (1993)]{Phillips93}
     {\sc Phillips, M. M.} 1993
     {\it Ap.J} {\bf 413}, L105.

\bibitem[Phillips \etal (1992)]{Phillips92}
     {\sc Phillips, M. M., Jacoby, G. H., Walker, A. R., Tonry, J. L., \& 
     Ciardullo, R.} 1992
     {\it B.A.A.S.} {\bf 180}, 1304.

\bibitem[Pierce \etal (1994)]{Pierce94}
     {\sc Pierce, M. J., Welch, D. L., McClure, R. D., van den Bergh, S.,
      Racine, R., \& Stetson, P. B.} 1994
     {\it Nature} {\bf 371}, 385--389.

\bibitem[Pottasch (1990)]{Pottasch90}
     {\sc Pottasch, S. R.} 1990
     {\it Astr.Ap.} {\bf 236}, 231.

\bibitem[Richer \& McCall (1992)]{RM92}
     {\sc Richer, M. G., \& McCall, M. L.} 1992
     {\it AJ} {\bf 103}, 54.

\bibitem[Richer \etal (1997)]{Richer97}
     {\sc Richer, M. G., McCall, M. L., \& Arimoto, N.} 1997
     {\it Astr.Ap.}, in press.

\bibitem[Riess \etal (1996)]{Riess96}
     {\sc Riess, A. G., Press, W. H., \& Kirshner, R. P.} 1996
     {\it Ap.J.}, submitted.

\bibitem[Sandage \& Tammann (1990)]{Sandage90}
     {\sc Sandage, A., \& Tammann, G. A.} 1990
     {\it Ap.J.} {\bf 365}, 1.

\bibitem[Sandage \etal (1996)]{Sandage96}
     {\sc Sandage, A., Saha, A., Tammann, G. A., Labhardt, L., Panagia, N.,
      \& Macchetto, F. D.} 1996
     {\it Ap.J.} {\bf 460}, L15--L18.

\bibitem[Simard \& Pritchet]{SP94}
     {\sc Simard, L., \& Pritchet, C. J.} 1994
     {\it AJ} {\bf 107}, 503.

\bibitem[Soffner \etal (1996)]{Soffner96}
     {\sc Soffner, T., M\'endez, R. H., Jacoby, G. H., Ciardullo, R., 
     Roth, M. M., \& Kudritzki, R. P.} 1996
     {\it Astr.Ap.} {\bf 306}, 9.

\bibitem[Stanghellini (1995)]{Stanghellini95} 
     {\sc Stanghellini, L.} 1995 
     {\it Ap.J.} {\bf 452}, 515-521.

\bibitem[Tammann (1992)]{Tammann92} 
     {\sc Tammann, G. A.} 1992 
     in {\it IAU Symposium 155, Planetary Nebulae} eds. R. Weinberger \& 
     A. Acker, 515.

\bibitem[Tanvir \etal (1995)]{Tanvir95}
     {\sc Tanvir, N. R., Shanks, T., Ferguson, H. C., \& Robinson, D. R. T.}
     1995
     {\it Nature} {\bf 377}, 27.

\bibitem[Vassiliadis \& Wood (1994)]{Vass}
     {\sc Vassiliadis, E., \& Wood, P.R.} 1994
     {\it Ap.J.Sup.} {\bf 92}, 125.

\bibitem[Weidemann (1987)]{Weidemann87}
     {\sc Weidemann, V.} 1987
     {\it Astr.Ap.} {\bf 188}, 74.

\bibitem[Weidemann \& Koester (1983)]{WK83}
     {\sc Weidemann, V., \& Koester, D.} 1983
     {\it Astr.Ap.} {\bf 121}, 7.

\bibitem[Wise \& Silva (1996)]{WS96}
     {\sc Wise, M. W., \& Silva, D. R.} 1996
     {\it Ap.J.} {\bf 461}. 155.

\bibitem[Witt \etal (1992)]{WTC92}
     {\sc Witt, A. N., Thronson, H. A., Jr., \& Capuano, J. M., Jr.} 1992
     {\it Ap.J.} {\bf 393}, 611.

\end{thebibliography}
\end{document}